\documentclass[aps,prd,twocolumn,showpacs,showkeys]{revtex4}
\usepackage{graphicx}
\usepackage{amsmath}
\usepackage{amssymb}
\usepackage{color}

\newcommand{\romane}{\mathrm{e}}
\newcommand{\romani}{\mathrm{i}}
\newcommand{\romand}{\mathrm{d}}

\newcommand{\fl}{\hspace*{-\mathindent}}

\usepackage[T2A]{fontenc}
\usepackage[utf8]{inputenc}
\usepackage[english,russian,ukrainian]{babel}

\begin{document}

\title[Density of states of Dirac-Landau levels]{Density of states of Dirac-Landau levels in a gapped graphene monolayer
under strain gradient}

\author{V.O. Shubnyi$^1$ and S.G. Sharapov$^2$}

\address{$^1$ Department of Physics,
Taras Shevchenko National University of Kiev,
6 Academician Glushkov ave.,
Kiev 03680, Ukraine}

\address{$^2$ Bogolyubov Institute for Theoretical Physics, National Academy of Science of Ukraine, 14-b
        Metrolohichna Street, Kiev 03680, Ukraine}

\email{sharapov@bitp.kiev.ua}
\vspace{10pt}


\begin{abstract}
\selectlanguage{english}
We study a gapped graphene monolayer in a combination of uniform
magnetic field and strain-induced uniform pseudomagnetic field.
The presence of two fields completely removes the valley degeneracy.
The resulting density of states shows a complicated behaviour that can be tuned by adjusting
the strength of the fields. We analyze how these features can be observed in the sublattice,
valley and full density of states. The analytical expression for the valley DOS is derived.
\end{abstract}

\selectlanguage{english}










\maketitle

\section{Introduction} \label{sec:intro}

The carbon atoms in monolayer graphene form  a honeycomb lattice due to
sp$^2$ hybridisation of their orbitals.
Since the honeycomb lattice is not a Bravais lattice, one has to consider the honeycomb lattice
as a triangular Bravais lattice with  two atoms per until cell.
Thus one naturally arrives at a two-component spinor wave function of the quasiparticle excitations in graphene
(see Ref. \cite{Mecklenburg2011PRL} for some analogies with a real spin).
These components reflect the amplitude of the electron wave function on the $A$ and $B$ sublattices.
The two-component form of the wave function along with the band structure results in the Dirac form of the
effective theory for graphene.

The Dirac fermions had shown up the celebrated magneto-transport  and STS
properties of graphene (see Refs.~\cite{CastroNeto2009RMP,Goerbig2011RMP,Andrei2012RPP} for the reviews).
Recently STM/STS measurements allowed not only to observe  relativistic Landau
levels, but also to resolve directly their sublattice specific features. By
resolving the density of states (DOS) on $A$ and $B$ sublattices of a gapped graphene, it was experimentally confirmed
\cite{Wang2015PRB} that the amplitude of the wave function of the lowest Landau
level (LLL) is unequally distributed between the sublattices depending on  its energy
sign.

In the presence of a gap $\Delta$ driven by  inversion symmetry breaking,
the LLL splits into two levels with the  energy $E_{0} =  \eta \Delta
\mbox{sgn} \, (eB) $,  where $\eta =\pm$ distinguishes  inequivalent $\mathbf{K}$ and
$\mathbf{K}^\prime$  points
of the Brillouin zone and an external magnetic field $\mathbf{B} = \mathbf{\nabla} \times \mathbf{A} = (0,0,B)$
is applied perpendicularly to  the plane of graphene along the positive $z$ axis \cite{Gusynin2007IJMPB}.
Here $e = -|e|$ is the electron charge and $\mathbf{A}$ is the vector electromagnetic potential.
The corresponding amplitudes of the wave function of the positive energy electron-like,
$0_{+}$, and negative energy hole-like, $0_{-}$, levels are  on $A$ and
$B$ sublattices. In other words, the individual sublattices are valley
polarized for the LLL~\cite{Settness2016PRB}.

An exciting opportunity for manipulating the amplitudes of the wave function on
the sublattices opens due to the close connection between the impact of
deformation and external electromagnetic field on the electronic structure of
graphene. Change in hopping energy between $A$ and $B$ atoms induced by strain
can be described by a vector potential $\mathbf{A}_{\mathrm{pm}}$ analogous to the
vector potential $\mathbf{A}$ of the external magnetic field (see Refs.~\cite{Vozmediano2010PR,Amorim2015PR} for a review).

The corresponding field, $\mathbf{B}_{\mathrm{pm}} = \mathbf{\nabla} \times \mathbf{A}_{\mathrm{pm}}$,
is called pseudomagnetic field (PMF), as it formally resembles the real magnetic field, with one crucial distinction that
it is directed oppositely in $\mathbf{K}$ and $\mathbf{K}^\prime$  valleys.
This implies that the LLL breaks the electron-hole symmetry, with the LLL energy, $E_{0} =  \Delta
\mbox{sgn} \, (B_{\mathrm{pm}}) $  for  both
$\mathbf{K}$ and $\mathbf{K}^\prime$ points.
Furthermore, the states corresponding to the LLL
are sublattice polarized, as they reside exclusively on either  $A$ or $B$ sublattice \cite{Settness2016PRB}.

While the formation of the LLL is associated with zero modes and does not
require a homogeneous PMF, to form higher Landau levels a uniform PMF is needed \cite{Roy2011PRB}.
This is in fact the main challenge \cite{Guinea2010NatPhys}  for the implementation of strained graphene,
although recently there has been some progress both in experiment
\cite{Levy2010Science,Klimov2012Science,Downs2016APL} and in theory~\cite{Zhu2015PRL}.

In the presence of either external magnetic field or deformation, the higher
energy levels from $\mathbf{K}$ and $\mathbf{K}^\prime$  points remain
degenerate. This degeneracy is lifted when both strain and magnetic field are
present. One of the interesting consequences of the lifting is that for
$|B_{\mathrm{pm}}| > | e B|$, the Hall conductivity is oscillating between $0$
and $\mp 2 e^2/h$~\cite{Roy2013PRB}.

The latest experiments \cite{Downs2016APL} show that it is possible to create a
homogeneous PMF of order of a few Tesla. Therefore, there is a good chance that
the STS/STM measurements of the Dirac-Landau levels similar to that done in
Ref.~\cite{Wang2015PRB} are now possible on strained graphene. Thus the purpose
of the present work is to study the DOS (including the sublattice resolved) in
a combination of a constant PMF $B_{\mathrm{pm}}$ created by non-uniform
strain and magnetic field $B$. In particular, we will look for the specific effects
related to the presence of nonzero gap $\Delta$ and lifting of the degeneracy between
$\mathbf{K}$ and $\mathbf{K}^\prime$ that can be observed in STS measurements.

The paper is organized as follows.  We begin by presenting in
section~\ref{sec:model} the model describing gapped monolayer graphene in the
combination of PMF and magnetic field. In section~\ref{sec:GF-DOS}  we provide
the definitions of the valley, sublattice and full DOS in terms of the Green's
function decomposed over Landau levels. The corresponding DOS are written in
section~\ref{sec:DOS-expressions} as the sums that in the case of the valley DOS
can be calculated analytically. The results for the DOS in the various regimes
are discussed in section~\ref{sec:results} and
conclusions are given in section~\ref{sec:concl}.

\section{Model}
\label{sec:model}

We consider gapped monolayer graphene in the continuum approximation described by the
effective Hamiltonian
\begin{equation} \label{Dirac-Hamiltonian} H = \left(
\begin{array}{cc}
H_{\mathbf{K}}  & 0 \\
0 & H_{\mathbf{K}^\prime}   \\
\end{array} \right),
\end{equation}
The full Hamiltonian  (\ref{Dirac-Hamiltonian}) acts on wave function with four components
 \begin{equation}
\Psi = \left(
         \begin{array}{c}
           \psi_{\mathbf{K}}^\bullet \\
           \psi_{\mathbf{K}}^\circ \\
           \psi_{\mathbf{K}^\prime}^\circ \\
           \psi_{\mathbf{K}^\prime}^\bullet \\
         \end{array}
       \right),
\end{equation}
where $\bullet$ and $\circ$ denote, respectively, $A$ and $B$ sublattices and
we followed the notations of Refs.~\cite{Gusynin2007IJMPB,Goerbig2011RMP} with
exchanging the sublattices in the  $\mathbf{K}^\prime$ valley. Thus the Hamiltonian  (\ref{Dirac-Hamiltonian})
includes two blocks corresponding to $\mathbf{K}$ and $\mathbf{K}^\prime$ valleys
\begin{eqnarray}
H_{\mathbf{K}} & = & v_F \pmb{\tau} \left(-i \hbar \mathbf{\nabla} - \frac{e}{c} \mathbf{A} - \mathbf{A}_{\mathrm{pm}} \right) + \tau_3 \Delta, \label{K1} \\
H_{\mathbf{K}^\prime} & = & -  v_F \pmb{\tau} \left(-i \hbar \mathbf{\nabla} - \frac{e}{c} \mathbf{A} + \mathbf{A}_{\mathrm{pm}} \right) - \tau_3 \Delta. \label{K2}
\end{eqnarray}
Here $\pmb{\tau} = (\tau_1, \tau_2)$ and $\tau_3$ are Pauli matrices
acting in the sublattice space,  $v_F$ is the Fermi velocity,
the gap $\Delta$ corresponds to the energy difference $2 \Delta$ between the $A$ and $B$ sublattices,
$\mathbf{A}$ and $\mathbf{A}_{\mathrm{pm}}$ are the
the electromagnetic and strain induced vector potentials, respectively.
We neglect the spin splitting, because for commonly used strengths of  magnetic
field the Zeeman splitting is small compared to the distance between the Landau levels.
For a fixed direction of  external magnetic field, the corresponding to $\mathbf{A}$ term in the
Hamiltonian breaks time-reversal symmetry, while the $\mathbf{A}_{\mathrm{pm}}$ term breaks the inversion symmetry
and leaves time-reversal symmetry unbroken.

With the $x$-axis aligned in the zigzag direction, the strain-induced vector potential reads \cite{Suzuura2002PRB,Ramezani2013SSC}
(see also the reviews
\cite{Vozmediano2010PR,Amorim2015PR})
\begin{equation}
\mathbf{A}_{\mathrm{pm}} = \frac{\hbar \beta \kappa }{2 a_0}  \left(
               \begin{array}{c}
                 u_{xx} - u_{yy} \\
                 -2 u_{xy} \\
               \end{array}
             \right),
\end{equation}
where $\beta = - \partial \ln t/\partial \ln a |_{a = a_0} \approx 3$ is the dimensionless
electron Gr\"{u}neisen parameter for the lattice deformation, $t$ the
nearest-neighbour hopping parameter, $\kappa \approx 1/3$ is a parameter related
to graphene's elastic property \cite{Suzuura2002PRB},
$a$ is the length of the
carbon-carbon bond, ($a_0$ is the length of the unstrained bond), and $u_{i j}$ with $i,j = x,y$ is
the strain tensor as defined in classical continuum
mechanics \cite{Vozmediano2010PR,Amorim2015PR}.
We also assume that the deformation is a pure shear, so that $u_{xx}+u_{yy} =0$, and there is no scalar potential term
in the Hamiltonian.

The sign of the PMF depends on the  valley, and, for example,
in $\mathbf{K}$ valley,
\begin{equation}
\label{PMF}
B_{\mathrm{pm}} = -\frac{\hbar \beta \kappa }{a_0} \left( \partial_x u _{xy} +
\frac{1}{2}\partial_y (u_{xx} - u_{yy})\right),
\end{equation}
whereas it has the opposite sign in $\mathbf{K}^\prime$ valley, because  $\mathbf{A}_{\mathrm{pm}}$
enters Eqs.~(\ref{K1}) and (\ref{K2}) with the opposite signs.
Eq.~(\ref{PMF}) illustrates the main problem in this field of research, viz. a uniform
PMF can only be created by a non-uniform strain
\cite{Guinea2010NatPhys}. As was already stated in the Introduction, considering the experimental progress
achieved in the field \cite{Downs2016APL}, we restrict ourselves to a constant PMF.
Thus we arrive at the model with two independent $\mathbf{K}$ points characterized by the following
combinations of the fields, $B_{\pm} = e B/c \pm B_{\mathrm{pm}}$.
A more complicated, but analytically  intractable case with a combination of a constant magnetic and inhomogeneous
pseudomagnetic fields was considered in Ref.~\cite{Kim2011PRB},
where a circularly symmetric strain is induced by a homogeneous load.

\section{Green's function, sublattice and valley resolved DOS}
\label{sec:GF-DOS}

Although it is straightforward to obtain the DOS directly from the solution of the
corresponding Dirac equation, we rely on the Green's function (GF) machinery that automatically
takes into account the degeneracy of levels and avoids the necessity to work with different directions of fields separately.
Since the  $\mathbf{K}$ points in the model (\ref{Dirac-Hamiltonian}) are independent,
we will use the GF's corresponding to the separate $\mathbf{K}$ points.
In particular, we are interested in the translation invariant part $\widetilde G$
of the GF that allows to derive both the DOS and the transport coefficients.
Its derivation using the Schwinger proper-time
method and decomposition over Landau-level poles has been
discussed in many papers (see, e.g., Refs. \cite{Chodos1990PRD,Gusynin1994PRL,Sharapov2003PRB,Miransky2015PR}).
Here we begin with the translation invariant part for $\mathbf{K}$ point written in the Matsubara
representation (we set $\hbar =c= k_B=1$ in what follows)
\begin{eqnarray} \label{GF}
 \fl   \widetilde G^{\mathbf{K}} (\romani \omega, \mathbf{p}) = \romane^{-\frac{p^2}{|B_+|}}
    \sum_{n = 0}^\infty (-1)^n
    \frac{G_n (B_+, \romani \omega, \mathbf{p})}
    {(\romani \omega)^2 - (M_n^{+})^2},\\
\fl    \omega = \pi(2m+1)T, \nonumber
\end{eqnarray}
where $T$ is the temperature,
\begin{equation}
M_n^{\pm} = \sqrt{\Delta^2 + 2 n v_F^2 |B_\pm|}
\end{equation}
are the energies of the relativistic Landau levels at
$\mathbf{K}$ and $\mathbf{K}^\prime $ points ($\eta = \pm$), respectively,
and the function
\begin{eqnarray}
\label{GF-num}
    \fl    G_n (B_+, \romani \omega, \vec p) & =  (\Delta\,\tau_3 +  \romani \omega)
     \left[ (1 + \tau_3 \, \mathrm{sgn}(B_+))L_n \left(\frac{2 p^2}{|B_+|}\right) \right. \nonumber \\
   & -  \left. (1 - \tau_3 \, \mathrm{sgn}(B_+))L_{n-1} \left(\frac{2 p^2}{|B_+|}\right)\right] \\
    &  - 4 v_F (p_x \tau_y + p_y \tau_x) L_{n-1}^{1}\left(\frac{2 p^2}{|B_+|}\right). \nonumber
\end{eqnarray}
Here  $L_n^\alpha(z)$ are the generalized Laguerre polynomials, and $L_n(z) \equiv L_n^0(z)$
($L_{-1}^1 \equiv 0$).
When deriving GF from the known wave-functions, the Laguerre polynomials originate from the integration
of two Hermite polynomials with proper weights. Looking at the structure of the GF (\ref{GF}), one can see
that the projectors $P_{\pm} =(1 \pm \tau_3 \mathrm{sgn}(B_+))/2$ take into account that, for example, for $B_+ > 0$,
the states on $A$ and $B$ sublattices involve $L_n$ and $L_{n-1}$, respectively.
The most general expression of the propagator in the presence of $B$, $B_{\mathrm{pm}}$ and various types
of the gaps is provided in \cite{Rybalka2015PRB}.

The corresponding contribution of the $\mathbf{K}$ point to the DOS per spin and unit area on $A$ and $B$ sublattices reads
\begin{equation}
\label{DOS-K}
D^{\mathbf{K}}_{A,B}(\epsilon) = \frac{1}{2\pi i} \int \frac{d^2 p}{(2 \pi)^2} [\widetilde G^{\mathbf{K}}_{ii}(\epsilon - i 0, \mathbf{p}) -
\widetilde G^{\mathbf{K}}_{ii}(\epsilon + i 0, \mathbf{p})]
\end{equation}
with $i=1,2$ for $A$ and $B$ sublattices, respectively. It follows from Eqs.~(\ref{K1}) and (\ref{K2})
that
\begin{equation} \label{GF-prime}
\widetilde G^{\mathbf{K}^\prime} (\romani \omega, \mathbf{p}) =
\widetilde G^{\mathbf{K}} (v_F \to - v_F, \Delta \to - \Delta, B_+ \to B_-)
\end{equation}
and
\begin{equation}
\label{DOS-K-prime}
D^{\mathbf{K}^\prime}_{B,A} (\epsilon)= \frac{1}{2\pi i} \int \frac{d^2 p}{(2 \pi)^2} [\widetilde G^{\mathbf{K}^\prime}_{ii}(\epsilon - i 0, \mathbf{p}) -
\widetilde G^{\mathbf{K}^\prime}_{ii}(\epsilon + i 0, \mathbf{p})]
\end{equation}
with $i=1,2$ for $B$ and $A$ sublattices, i.e. exchanging the sublattices.
While the valley resolved DOS  presents a theoretical interest and will also be considered below,
the STS measurements allow to observe the
full DOS involving two valleys on each sublattice
\begin{equation}
\label{AB-DOS}
D_{A,B} (\epsilon)= D^{\mathbf{K}}_{A,B} (\epsilon) + D^{\mathbf{K}^\prime}_{A,B} (\epsilon).
\end{equation}
We will also consider the valley resolved but summed over sublattices DOS
\begin{equation}
\label{valley-DOS}
D^{\mathbf{K}, \mathbf{K}^\prime} (\epsilon) = D^{\mathbf{K}, \mathbf{K}^\prime}_{A} (\epsilon) +
D^{\mathbf{K}, \mathbf{K}^\prime}_{B} (\epsilon)
\end{equation}
which presents interest for valleytronics. Finally, the full DOS can also be found by summing the
valley resolved DOS
\begin{equation}
\label{DOS}
D (\epsilon)= D^{\mathbf{K}}(\epsilon) + D^{\mathbf{K}^\prime}(\epsilon).
\end{equation}

\section{Expressions for numerical and analytical calculation of the DOS}
\label{sec:DOS-expressions}

Using the integral \cite{Gradstein:book}
\begin{equation}
\int \frac{\romand^2 p}{(2 \pi)^2} L_n \left( \frac{2 p^2}{|B|} \right) \exp \left(- \frac{p^2}{|B|} \right) =
\frac{(-1)^n}{2 \pi} \frac{|B|}{2}
\end{equation}
and evaluating the discontinuity of the GF we arrive at the final result
\begin{equation}
D^{\mathbf{K},\mathbf{K}^\prime}_{A,B} (\epsilon)= D^{\mathbf{K},\mathbf{K}^\prime (0)}_{A,B} (\epsilon) +
D^{\mathbf{K},\mathbf{K}^\prime (n \geq 1)}_{A,B} (\epsilon).
\end{equation}
Here  $D^{\mathbf{K},\mathbf{K}^\prime (0)}_{A,B} (\epsilon)$ is
the LLL contribution to the valley and sublattice resolved DOS and
$D^{\mathbf{K},\mathbf{K}^\prime (n \geq 1)}_{A,B} (\epsilon)$
is the corresponding contribution from the Landau levels with $n \geq1$.
Explicit expressions for these terms are
\begin{eqnarray}
\label{A-DOS-LLL}
\fl
    & D^{\mathbf{K} (0)}_{A} (\epsilon) = \frac{|B_+|}{2 \pi} \theta (B_+) \delta(\epsilon - \Delta),
\nonumber\\
    & D^{\mathbf{K}^\prime (0)}_{A} (\epsilon) = \frac{|B_-|}{2 \pi} \theta (-B_-) \delta(\epsilon - \Delta),
\end{eqnarray}
\begin{eqnarray}
\label{B-DOS-LLL}
\fl
    & D^{\mathbf{K} (0)}_{B} (\epsilon) = \frac{|B_+|}{2 \pi} \theta (-B_+) \delta(\epsilon + \Delta),
\nonumber\\
    & D^{\mathbf{K}^\prime (0)}_{B} (\epsilon) = \frac{|B_-|}{2 \pi} \theta (B_-) \delta(\epsilon + \Delta),
\end{eqnarray}
and for the Landau levels with $n \geq1$,
\begin{eqnarray}
\label{AK-DOS-n>1}
    \fl
     D^{\mathbf{K}, \mathbf{K}^\prime (n\geq1)}_{A} (\epsilon) =
     \sum_{n = 1}^\infty &&
    \frac{|B_\pm|}{2\pi}  \Big[
         \frac{M_n^\pm + \Delta}{2 M_n^\pm}
    \delta(\epsilon - M_n^\pm)
    \nonumber\\
    &&+ \frac{M_n^\pm - \Delta}{2 M_n^\pm}
    \delta(\epsilon + M_n^\pm)    \Big ],
\end{eqnarray}
\begin{eqnarray}
\label{BK-DOS-n>1}
    \fl
D^{\mathbf{K}, \mathbf{K}^\prime (n\geq1)}_{B} (\epsilon) =
    \sum_{n = 1}^\infty &&
    \frac{|B_\pm|}{2\pi} \Big[
    \frac{M_n^\pm - \Delta}{2 M_n^\pm}
    \delta(\epsilon - M_n^\pm)
    \nonumber\\
    && + \frac{M_n^\pm + \Delta}{2 M_n^\pm}
    \delta(\epsilon + M_n^\pm)    \Big],
\end{eqnarray}
where $\pm$ sign corresponds to $\mathbf{K}$ and $\mathbf{K}^\prime$ points.
As expected,  presence of PMF removes  degeneracy of the levels with $n \geq 1$ \cite{Roy2013PRB} .

In section~\ref{sec:results} we compute the sublattice and valley resolved DOS numerically on the base of
Eqs.~(\ref{A-DOS-LLL}), (\ref{B-DOS-LLL}), (\ref{AK-DOS-n>1}) and (\ref{BK-DOS-n>1})
by widening $\delta$-fuction peaks to
a Lorentzian shape, viz.
\begin{equation}
\label{widening}
\delta(\epsilon-M_n) \to \frac{1}{\pi}  \frac{1}{(\epsilon - M_n)^2 + \Gamma_n^2},
\end{equation}
where $\Gamma_n$ is the $n$-th level width.
Such  broadening of Landau levels with a constant $\Gamma$
was found to be rather a good approximation valid in not
very strong magnetic fields.

\subsection{The DOS in the zero pseudomagnetic field}

Setting $B_{\mathrm{pm}}=0$ we  recover the well-known  results
that were experimentally observed in \cite{Wang2015PRB}. Then Eqs.~(\ref{A-DOS-LLL})
and (\ref{B-DOS-LLL}) result in the sublattice DOS
\begin{equation}
\fl D^{(0)}_{A} (\epsilon) = \frac{|eB|}{2 \pi} \delta(\epsilon - \Delta ), \quad
 D^{(0)}_{B} (\epsilon) = \frac{|eB|}{2 \pi} \delta(\epsilon + \Delta ).
\end{equation}
This confirms that the LLL is valley polarized, because
each LLL contribution to the DOS comes from either $\mathbf{K}$ or $\mathbf{K}^\prime$ valley,
as discussed in the Introduction.
This feature has to be contrasted with the valley resolved but summed over the two sublattices DOS
\begin{eqnarray}
\fl
    D^{\mathbf{K},\mathbf{K}^\prime (0)} (\epsilon) = && D^{\mathbf{K},\mathbf{K}^\prime (0)}_{A} (\epsilon) +
D^{\mathbf{K},\mathbf{K}^\prime (0)}_{B} (\epsilon)
    \nonumber\\
    = && \frac{|eB|}{2 \pi} \delta(\epsilon - \eta \Delta \mbox{sgn} \, (e B) ).
\end{eqnarray}
For $n\geq 1$ the levels at $\mathbf{K}$ and $\mathbf{K}^\prime$ points described by
Eqs.~(\ref{AK-DOS-n>1}) and (\ref{BK-DOS-n>1}) are degenerate, but the DOS on $A$ and $B$
sublattices differs and this effect is observable  \cite{Wang2015PRB}.

\subsection{The DOS in the zero magnetic field}

Setting $eB =0$ we obtain from Eqs.~(\ref{A-DOS-LLL}) and (\ref{B-DOS-LLL}) that the LLL contribution to
the sublattice DOS is
\begin{eqnarray}
     D^{(0)}_{A} (\epsilon) &=&
\frac{|B_{\mathrm{pm}}|}{2\pi} \theta(B_{\mathrm{pm}})\delta(\epsilon - \Delta),
\nonumber \\
     D^{(0)}_{B} (\epsilon) &=&
\frac{|B_{\mathrm{pm}}|}{2\pi} \theta(-B_{\mathrm{pm}})\delta(\epsilon + \Delta).
\end{eqnarray}
This confirms that the LLL is sublattice polarized, as discussed in the Introduction.

\subsection{Analytical expression for the valley DOS}

Although the expressions for the sublattice and valley DOS presented in Sec.~\ref{sec:GF-DOS}
are sufficient for the numerical study presented in Sec.~\ref{sec:results}, it is always useful to have  a simple
analytical expression for the DOS. One can notice that the  the valley DOS, Eq.~(\ref{valley-DOS})
is the sum of delta-functions (or Lorentzians when the the level widening
is taken into account), because  the sum of the weight factors $(M_n^{\pm} \pm \Delta)/(2M_n^{\pm}$) present in
Eqs.~(\ref{AK-DOS-n>1}) and (\ref{BK-DOS-n>1}) gives $1$. This allows one to use
the results of Ref.~\cite{Sharapov2004PRB}, and calculate the sum over Landau levels analytically
\begin{eqnarray}
\label{valley-DOS-anal}
       D^{\mathbf{K}, \mathbf{K}^\prime}(\epsilon) =&& \frac{1}{2 \pi^2}  \Bigg \{
-|B_{\pm}| \theta(\mp B_{\pm}){\Gamma \over (\epsilon-\Delta)^2 + \Gamma^2}
\nonumber\\
       && -|B_{\pm}| \theta(\pm B_{\pm}){\Gamma \over (\epsilon+\Delta)^2 + \Gamma^2}
\nonumber \\
      &&   + \Gamma \ln {\Lambda^2 \over 2 |B_{\pm}|}
\\
      && - \mathrm{Im}  \left[ (\epsilon + \romani \Gamma)
    \psi \biggl( {\Delta^2-(\epsilon + \romani \Gamma)^2
    \over 2 |B_{\pm}|} \biggr)  \right]
        \Bigg\}.
\nonumber
\end{eqnarray}
Here $\psi$ is the digamma function, $\pm$ sign corresponds to $\mathbf{K}$ and $\mathbf{K}^\prime$ points, the width of all levels
$\Gamma$ is assumed to be the same,  and $\Lambda$ is the cutoff energy that has the order of bandwidth.
Eq.~(\ref{valley-DOS-anal}) differs from Eq.~(4.15) of Ref.~\cite{Sharapov2004PRB} by the first two terms. In the present
case they take care of the electron hole asymmetry of the LLL, while in \cite{Sharapov2004PRB} both
$\mathbf{K}$ and $\mathbf{K}^\prime$ points contribute to the full DOS.
The advantage of Eq.~(\ref{valley-DOS-anal}) is that it allows to consider the low field regime when the direct numerical
summation over many Landau levels is consuming.

\section{Results }
\label{sec:results}

Now we use Eqs.~(\ref{A-DOS-LLL}), (\ref{B-DOS-LLL}), (\ref{AK-DOS-n>1}) and (\ref{BK-DOS-n>1})
to study the valley (\ref{valley-DOS}), sublattice (\ref{AB-DOS}) and the full
(\ref{DOS}) DOS numerically.
For simplicity we assume that all Landau levels have the same width $\Gamma$. In this case
the valley DOS and then the full DOS can also be calculated using Eq.~(\ref{valley-DOS-anal}).
To fit real experimental data \cite{Ponomarenko2010PRL} it may be necessary to consider the width, $\Gamma_n$, dependent on the Landau
level index. This can be easily done in the framework of numerical computation of
the sum over Landau levels. However, when all levels have the same width and one is interested in the
valley DOS, it is more efficient to compute it from the analytical expression (\ref{valley-DOS-anal})
which is easier to use in the low field regime. In all numerical work we take the value of the Fermi velocity,
$v_F = 10^6 \, \mbox{m/s}$ that corresponds to the Landau energy scale,
${\epsilon_0 =(\hbar v_F^2 B_{\pm})^{1/2}} =25.7\sqrt{B_{\pm}{\rm [T]}}\ {\rm meV}$.
The gap $\Delta$ that lifts the energy degeneracy of the $A$ and $B$ sublattices and breaks the inversion symmetry
was observed for a graphene monolayer on top of SiC, graphite \cite{Andrei2012RPP}, and hexagonal boron nitride \cite{Gorbachev2014Science}.
Its value ranges from 10 meV to several tens of meV.

\begin{figure}[!h]
\raggedleft{
\includegraphics[width=1.\linewidth]{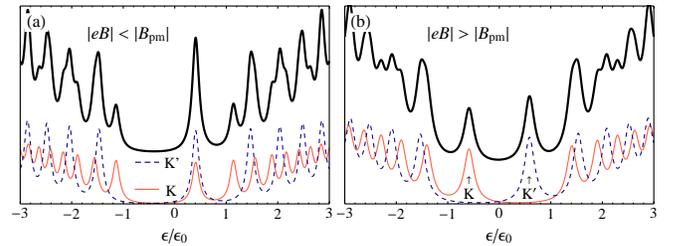}}
\caption{(Colour online) The full DOS, $D (\epsilon)$, (thick solid) and the valley-resolved
    DOS, $D^{\mathbf{K}, \mathbf{K}^\prime} (\epsilon)$, (thin solid and thin dashed) in arbitrary units as the functions
    of energy $\epsilon/\epsilon_0$, where  ${\epsilon_0 =(\hbar v_F^2 B_+)^{1/2}}
    =25.7\sqrt{B_{+}{\rm [T]}}\ {\rm meV}$ is the Landau scale for $\mathbf{K}$ valley.
    Left panel: (a): the fields $B = 5\ {\rm T}$ and $B_{\mathrm{pm}} = 18\ {\rm T}$.
    Right panel: (b): the fields  $B = 10\ {\rm T}$ and $B_{\mathrm{pm}} = 1\ {\rm T}$.
    The gap $\Delta = 50\ {\rm meV}$ and the scattering rate $\Gamma = 10\ {\rm meV}$ in the both cases.
    } \label{fig:1}
\end{figure}

Fig.~\ref{fig:1} demonstrates how the full density of states, Eq.~(\ref{DOS}), is
formed by the contributions from the valley resolved DOS, Eq.~(\ref{valley-DOS}): left panel (a) is for $|eB| < |B_{\mathrm{pm}}|$
and the right panel (b) is for $|eB| > |B_{\mathrm{pm}}|$.
The two curves (thin solid red and thin dashed blue) in the bottom part of the figure show the valley resolved DOS, $D^{\mathbf{K}, \mathbf{K}^\prime} (\epsilon)$. The curves for $\mathbf{K}$ and $\mathbf{K}^\prime$ points
have the peaks corresponding to the relativistic Landau levels with the energies $\sim \pm \sqrt{n |B_{\pm}|}$.
The positions of the peaks corresponding to the LLL with $E_0 = \pm \Delta$ distinguish
the cases (a) the PMF dominated regime, $|eB| < |B_{\mathrm{pm}}|$, when both peaks have the same sign of the energy, and (b)
the magnetic field dominated regime, $|eB| > |B_{\mathrm{pm}}|$, when the peaks have the opposite energy sign.
We checked that the same curves also follow from the analytical expression (\ref{valley-DOS-anal}).
Those are rather trivial consequences of having a superposition of magnetic and PMF.

The full DOS $D(\epsilon)$ shown by thick black curve in the two panels obviously
has  two series of peaks.
One could see that in the special cases, the difference
between two curves is substantial, and resulting DOS curve has irregular
features and/or masked peaks.  Depending on the values of the effective fields $B_{\pm}$ , the Landau levels could
be viewed as a splitting of one level (in case $|B_{+}| \approx |B_{-}|$) or
as the two largely independent series, as for the case shown in Fig.~\ref{fig:1}.

\begin{figure}[!h]
\centering{
\includegraphics[width=1.\linewidth]{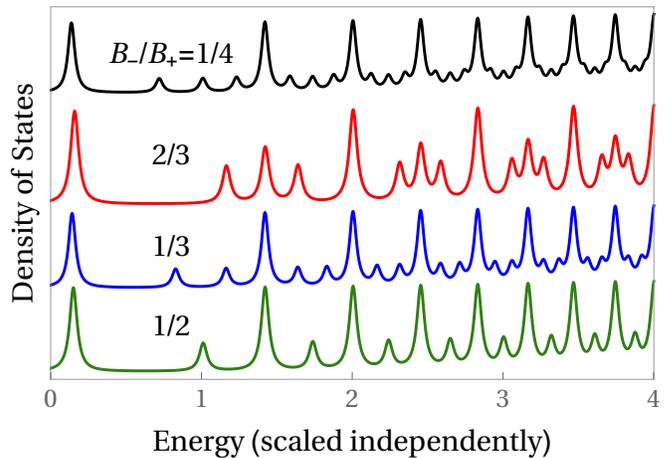}}
\caption{(Colour online)  The full DOS, $D (\epsilon)$,
in arbitrary units as the functions of energy $\epsilon/\epsilon_0$, where  ${\epsilon_0 =(\hbar v_F^2 B_+)^{1/2}}
=25.7\sqrt{B_{+}{\rm [T]}}\ {\rm meV}$. The green curve is for $B_-/B_+ =1/2$,
the blue curve is for  $B_-/B_+ =1/3$, the red curve is for $B_-/B_+ =2/3$ and
the black curve is for $B_-/B_+ =1/4$. The gap $\Delta = 10\ {\rm meV}$ and the scattering rate $\Gamma = 2\ {\rm meV}$.
} \label{fig:2}
\end{figure}

Let us look closer at the pattern that overlapping Landau levels
may create for certain values of $B$ and $B_{\mathrm{pm}}$.
The energies of the Landau levels with indices $n_+$, $n_-$ for $\mathbf{K}$ and $\mathbf{K}^\prime$ points
coincide, viz. $M_{n_+} = M_{n_-}$ if there exist some values of $B_+$ and $B_-$
satisfying the condition, $|B_+| n_+ = |B_-| n_-$. This implies that the fraction
$|B_+|/|B_-| = a/b$ has to be rational.
In terms of the initial fields $B$ and $B_{\mathrm{pm}}$  this condition implies that
\begin{equation}
    e B = \frac{1 - a/b}{1 + a/b} B_{\mathrm{pm}}.
\end{equation}
The corresponding beating patterns for four values of the fraction $a/b$ are shown in Fig.~\ref{fig:2}.
The lowest (green) curve is for the simplest case, $B_-/B_+ =1/2$.
Each second level with coinciding energies is enhanced.
The second from the bottom (blue) curve is for  $B_-/B_+ =1/3$. In this case an enhancement
occurs for each third level.
The third from the bottom (red) curve is for  $B_-/B_+ =2/3$ has even more tricky pattern with
the highest each third level and each forth level of an intermediate height. The curve on the top
(black) is for $B_-/B_+ =1/4$.

\begin{figure}[!h]
\centering{
\includegraphics[width=1.\linewidth]{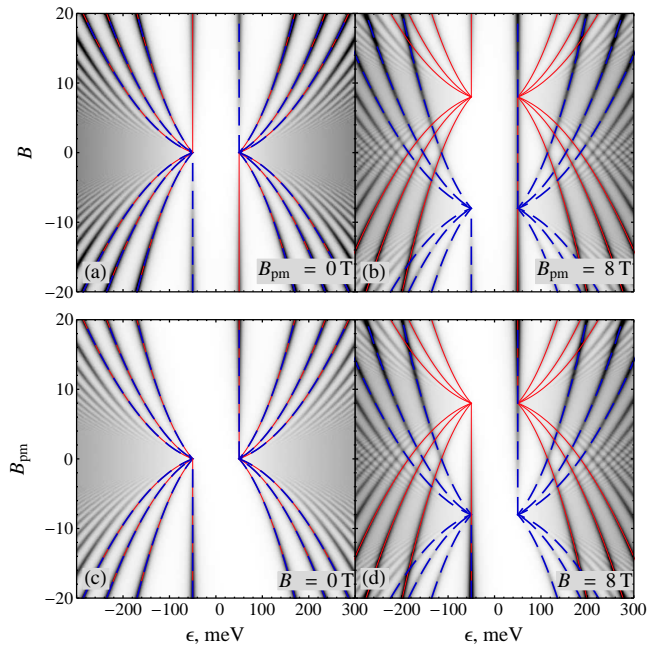}}
\caption{Density map of the full DOS $D (\epsilon, B,B_{\mathrm{pm}})$
as a function of energy $\epsilon$ in ${\rm meV}$, magnetic field $B$ in ${\rm T}$
and PMF in ${\rm T}$.
Top left panel: (a): for a constant $B_{\mathrm{pm}} =  0 \ {\rm T}$.
Top right panel: (b): for a constant $B_{\mathrm{pm}} =  8\ {\rm T}$.
Bottom left panel: (c): for a constant $B = 0\ {\rm T}$.
Bottom right panel: (d): for a constant $B =  8 \ {\rm T}$.
The density-map is overlaid with red and blue curves that  show the
position of the peaks originating from $\mathbf{K}$ and $\mathbf{K}^\prime$ points.
The gap $\Delta = 50\ {\rm meV}$, the scattering rate $\Gamma = 5\ {\rm meV}$
and the Landau scale ${\epsilon_0 =(\hbar v_F^2 B_{\pm})^{1/2}}
=25.7\sqrt{B_{\pm}{\rm [T]}}\ {\rm meV}$
in all cases.}
    \label{fig:3}
\end{figure}

It is instructive to represent the dependences of the full DOS, $D (\epsilon)$,
on the fields $B$ and $B_{\mathrm{pm}}$ employing the density plot.
Since a wide range of the fields is involved, its consideration in the low field regime
may demand summation over many Landau levels. Thus we use  Eq.~(\ref{valley-DOS-anal}), where
the summation is done analytically.
Figs.~\ref{fig:3}~(a) and (b) on the top panel show the full DOS, $D (\epsilon, B,B_{\mathrm{pm}})$
as a function of energy $\epsilon$ in ${\rm meV}$ and magnetic field $B$ in ${\rm T}$
for $B_{\mathrm{pm}} = 0\ {\rm T}$ and $B_{\mathrm{pm}} = 8\ {\rm T}$.
Figs.~\ref{fig:3} (c) and (d) in the bottom panel  show the full DOS, $D (\epsilon, B, B_{\mathrm{pm}})$
as a function of energy $\epsilon$ in ${\rm meV}$ and PMF $B_{\mathrm{pm}}$ in ${\rm T}$
for $B = 0\ {\rm T}$ and $B = 8\ {\rm T}$.
The density plot is partly overlaid with the solid (red)  and dashed (blue) curves that show position of
the peaks in the DOS originating from the Landau levels at
$\mathbf{K}$ and $\mathbf{K}^\prime$ points, respectively.
Fig.~\ref{fig:3}~(a) (top left panel) describes unstrained graphene. The Landau levels fan away
from the Dirac point at $\epsilon =0$. One can find a similar DOS map for
the STS measurements \cite{Andrei2012RPP} of graphene on chlorinated SiO$_2$.
In the real case the spectra are distorted  at low fields  due to the substrate induced
disorder and are strongly position dependent. The density plot \cite{Andrei2012RPP}  allows
to observe at higher fields the sequence of broadened Landau levels with separated peaks.
Fig.~\ref{fig:3}~(c) (bottom left panel) describes strained graphene in zero magnetic field. It is almost identical
to Fig.~\ref{fig:3}~(a)
except to the LLL that in the case of strained graphene breaks the electron-hole symmetry.
Fig.~\ref{fig:3}~(b) and (d) (right top and bottom panels) describe strained graphene in the external magnetic field.
This case was also studied experimentally in \cite{Klimov2012Science}, where SMT and STS measurements were made
on the deformed by gating graphene drumhead.

Comparing all these panels we observe that in the presence of both PMF and magnetic field
there exist regions of intersecting Landau levels with the opposite slope that are related to the opposite valleys.
In Fig.~\ref{fig:3}~(b) this is the region with $ |eB| < |B_{\mathrm{pm}}|  $, while
Fig.~\ref{fig:3}~(d) the corresponding region is seen for $ |B_{\mathrm{pm}}| < |e B|  $.
This behaviour of Landau levels is almost obvious in the presented case.
However, in the case of poorly resolved Landau levels this feature can be rather helpful
for proving the presence of both PMF and magnetic field.

\begin{figure}[!h]
    \centering{
\includegraphics[width=\linewidth]{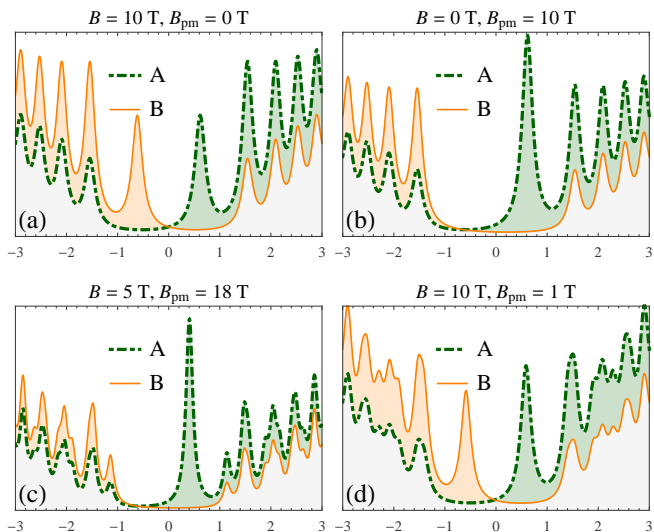}}
\caption{(Colour online)
The sublattice resolved DOS, $D_A (\epsilon)$ (dashed) and $D_B (\epsilon)$ (solid)
in arbitrary units as the functions of energy $\epsilon/\epsilon_0$.
Top left panel: (a): in the absence of PMF, $B_{\mathrm{pm}} =0$ and $B = 10\ {\rm T}$.
Top right panel: (b): in the absence of magnetic   $B = 0\ {\rm T}$, $B_{\mathrm{pm}} = 10\ {\rm T}$.
Bottom left panel: (c): the fields $B = 5\ {\rm T}$ and $B_{\mathrm{pm}} = 18\ {\rm T}$.
Bottom left panel:  (d) the fields  $B = 10\ {\rm T}$ and $B_{\mathrm{pm}} = 1\ {\rm T}$.
The gap $\Delta = 50\ {\rm meV}$ and the scattering rate $\Gamma = 10\ {\rm meV}$ in all cases.
        }
    \label{fig:4}
\end{figure}

Finally, we illustrate in Fig.~\ref{fig:4} how the full DOS is
distributed between the sublattices.
The DOS on $A$ and $B$ sublattices, $D_{A,B} (\epsilon)$, are shown by (dashed) green
and  solid (orange) curves, respectively.
Fig.~\ref{fig:4}~(a) (top left panel) describes unstrained graphene in the external magnetic field.
It corresponds to the situation studied experimentally in \cite{Wang2015PRB}. We observe that
the positive (negative) energy states reside on $A$ ($B$) sublattice. Since these states are associated with  different
valleys, the LLL  is indeed  valley polarized. Furthermore, the sublattice asymmetry is also seen for
higher levels, because we took a large value of the gap $\Delta = 50\ {\rm meV}$.
Fig.~\ref{fig:4}~(b) (top right panel) describes  strained graphene in zero magnetic field.
As it should be, the LLL is indeed completely sublattice polarized, while higher levels are polarized in the same fashion as
in Fig.~\ref{fig:4}~(a).
The PMF dominated regime, $|eB| < |B_{\mathrm{pm}}|$, is shown in Fig.~\ref{fig:4}~(c) (bottom left panel). The LLL polarization
is similar with Fig.~\ref{fig:4}~(b). The magnetic field dominated regime,   $|eB| > |B_{\mathrm{pm}}|$, is shown
in Fig.~\ref{fig:4}~(d) (bottom right panel)  and it is similar to Fig.~\ref{fig:4}~(a).
Fig.~\ref{fig:4}~(c) and (d) are computed for the
same values of the parameters as Fig.~\ref{fig:1}~(a) and (b), respectively.
When both fields are present the asymmetry between the sublattices can be enhanced even for higher levels.

We note that in the present work the sublattice asymmetry is directly brought by the inversion symmetry
gap $\Delta$. We established that the presence of PMF and magnetic field further enhances this effect.
It is shown in \cite{Schneider2015PRB} that the local sublattice symmetry can be broken just by the deformation.
This deformation  is not a pure shear, so it produces not only the PMF, but also a scalar potential.

\section{Conclusion} \label{sec:concl}

In the present work we had in mind that the sublattice resolved DOS can be measured
by STS. However,  the full DOS can also be experimentally found by measuring the
quantum capacitance \cite{Ponomarenko2010PRL} which is proportional to
the thermally smeared DOS. The corresponding convolution with a Fermi distribution
is easily expressed in terms of the digamma function  \cite{Gusynin2014FNT},
so that the presented here results can be easily applied for this case.

In conclusion we note, that controlling the valley degree of freedom is important
for possible valleytronics applications of the new materials. In this respect a simultaneous tuning
of the strain (PMF) and magnetic field is rather useful, because it allows to remove
the valley degeneracy. Thus the experimental testing of the features discussed in this work
would be helpful for development of valleytronics.


We gratefully acknowledge E.V.~Gorbar, V.P. Gusynin and V.M. Loktev for helpful discussions.
S.G.Sh. acknowledges the the support from the Ukrainian State Grant for Fundamental
Research No. 0117U00236
and the support of EC for the RISE Project CoExAN GA644076.


\end{document}